\documentclass[11pt]{article}
\usepackage[margin=1in]{geometry} 
\usepackage{graphicx, float, appendix, booktabs, hyperref}
\usepackage{amssymb, amsmath, amsthm, bm, bbm}
\usepackage{natbib}

\hypersetup{
  colorlinks,
  linkcolor = {blue},
  citecolor = {blue},
  urlcolor = {blue}
}

\newcommand{\phref}{\hat{\tau}^2_\text{ref}}
\newcommand{\phrefk}[1][]{\hat{\tau}^2_{\text{ref}}(#1)}
\newcommand{\phhyb}{\hat{\tau}^2_\text{hyb}}
\newcommand{\deffh}{\hat{d}_{\text{eff}}}
\newcommand{\deff}{d_{\text{eff}}}
\DeclareMathOperator*{\argmin}{arg\,min}

\begin{document}

\title{A note on the amount of information borrowed from external data in hybrid controlled trials with time-to-event outcomes}

\author{Brian D. Segal and W. Katherine Tan\\
Flatiron Health}

\maketitle

\begin{abstract}
In situations where it is difficult to enroll patients in randomized controlled trials, external data can improve efficiency and feasibility. In such cases, adaptive trial designs could be used to decrease enrollment in the control arm of the trial by updating the randomization ratio at the interim analysis. Updating the randomization ratio requires an estimate of the amount of information effectively borrowed from external data, which is typically done with a linear approximation. However, this linear approximation is not always a reliable estimate, which could potentially lead to sub-optimal randomization ratio updates. In this note, we highlight this issue through simulations for exponential time-to-event outcomes, because in this simple setting there is an exact solution available for comparison. We also propose a potential generalization that could complement the linear approximation in more complex settings, discuss challenges for this generalization, and recommend best practices for computing and interpreting estimates of the effective number of events borrowed.
\end{abstract}

\section{Introduction}

In situations where it is difficult to enroll patients in randomized controlled trials, such as in rare diseases, external data can be used to improve efficiency and feasibility. For example, it may be possible to borrow data from historical trials meeting Pocock's criteria \citep{pocock1976}. In some circumstances it may also be possible to borrow data derived from real-world sources, such as electronic health records \citep{thomas2019}. We use the term \textit{hybrid controlled trials} to refer to randomized controlled trials in which the trial control arm is augmented with external data. There are several approaches for borrowing data from external comparators, ranging from simple test-then-pool decision rules to more complex Bayesian methods \citep{chen2000, ibrahim2000, duan2006,neuenschwander2009, hobbs2012, viele2014, ibrahim2015, van2017}.

\citet{hobbs2013} proposed an adaptive trial design in which the randomization ratio is updated at the interim analysis in the absence of strong evidence for outcome heterogeneity between randomized controls and external comparators. Updating the randomization ratio requires an estimate of the amount of information effectively borrowed from external data, and \citet{hobbs2013} proposed a linear approximation referred to as the effective historical sample size (EHSS) which has also been used in subsequent work \citep{chen2018, lewis2019, wu2019, normington2020}. Based on our simulation results, this linear approximation appears reasonable in many settings but can have large bias and variance in others, potentially leading to suboptimal randomization ratio updates.

In this note, we highlight this issue through simulations for exponential time-to-event outcomes, because in this simple setting there is an exact solution available for comparison. We also propose a potential generalization similar to the approach of \citet{morita2008, morita2012} that could complement the linear approximation of \citet{hobbs2013} in more complex settings, discuss challenges for this generalization, and recommend best practices for computing and interpreting estimates of the effective number of events borrowed.

\section{Estimating the amount of external information borrowed}

Borrowing data from external data can increase the precision of estimated treatment effects. Let $\phref$ be the precision of the log hazard ratio (HR) between the experimental and control arms at the interim when excluding external data (the reference model) and let $\phhyb$ be the precision at the interim when including external data (the hybrid model). Also, let $n_C$ and $n_E$ be the number of patients randomized to the control and experimental arms, respectively, by the time of the interim analysis. \citet{hobbs2013} proposed to approximate the EHSS as
\begin{equation}
  \text{EHSS} \approx (n_C + n_E) \left( \frac{\phhyb}{\phref} - 1 \right).
  \label{ehss}
\end{equation}
The approximation in (\ref{ehss}) aims to measure the number of additional patients that would need to be randomized to the control arm in order to achieve the same precision increase obtained by borrowing from external data.

In what follows, we make two shifts: 1) focusing on the number of events instead of number of patients, because in time-to-event analyses precision is primarily a function of the number of events, and 2) generalizing terminology by referring to `effective number of external events', as opposed to `historical trial events', because borrowing may be possible from real-world as well as historical trial data.

Focusing on the number of events allows for a more direct development, but randomization ratio changes are based on the number of patients effectively borrowed. Therefore, we must convert to the effective external sample size. To do so, we would need to make assumptions about the number of additional patients required in order to observe the additional number of events. For example, let $\kappa$ be the proportion of patients enrolled in the trial control arm who have experienced an event by the interim analysis. Assuming the additional patients would have been enrolled at the same rate and over the same amount of enrollment time and study follow-up time as patients in the trial control arm, we could estimate the effective number of external patients by dividing the effective number of external events ($\deffh$ below) by $\kappa$, i.e. $\deffh / \kappa$.

\subsection{Exact solution for exponential data}

Let $d_C$ and $d_E$ be the number of events in the control and experimental arm, respectively, of the trial at the interim analysis. For unadjusted exponential models, the precision of the log HR is \citep{kalbfleisch2002}
\begin{align}
  \phref = \frac{d_C d_E}{(d_C + d_E)}.
  \label{exp_prec}
\end{align}
We denote the estimated number of events effectively borrowed from the external comparator at the interim analysis as $\deffh$. The hybrid model effectively has $d_C + \deffh$ events in the control arm at interim. Substituting $d_C + \deffh$ for $d_C$ in (\ref{exp_prec}), we get
\begin{align}
  \phhyb = \frac{(d_C + \deffh) d_E}{d_C + d_E + \deffh}.
  \label{exp_prec_hyb}
\end{align}
Solving (\ref{exp_prec_hyb}) for $\deffh$, we get
\begin{equation}
   \deffh = \frac{\phhyb (d_C + d_E) - d_C d_E}{d_E - \phhyb}.
  \label{ENEE_exact}
\end{equation}

When using unadjusted exponential models to analyze trial data, (\ref{ENEE_exact}) gives the exact number of additional events that the trial control arm would have needed by the interim analysis in order to achieve the same precision as the model that borrows from the external data. This assumes that the trial analysis is done with an unadjusted model, but allows for arbitrary borrowing models, including ones with covariate adjustments and/or weighting.

\subsection{Generalization to non-exponential models \label{sec:gen}}

In unadjusted exponential models, precision is purely a function of the number of events that have occurred. This makes it straightforward to estimate the effective number of external events. With other commonly used outcome models, including Cox regression, precision is not solely a function of the number of events. Consequently, there is not always a unique generalizable solution to $\deffh$.

However, it is still possible to define a generalization of (\ref{ENEE_exact}) that can typically be solved under certain assumptions. Similar to \citet{morita2008, morita2012} we define the effective number of external events as
\begin{equation}
  \deffh := \argmin_{\deff} | \phrefk[\deff] - \phhyb |
  \label{enee_general}
\end{equation}
where $\phrefk[\deff]$ is the hypothetical log HR precision that would have been obtained if $\deff$ additional events had occurred in the control arm via additional trial enrollment and without external data. We note that (\ref{ENEE_exact}) is the exact solution to (\ref{enee_general}) in the case of unadjusted exponential models.

To solve (\ref{enee_general}), we assume that the additional controls would have had the same distribution of covariates, enrollment times, and events as the actual trial controls. Under these assumptions, we can apply a common case weight $w$ to all trial control patients, which effectively multiplies the number of controls, and thus the number of control arm events, by $w$. For a given case weight $w$, the effective number of control events in the weighted trial dataset is $w d_C$, resulting in $\deffh = w d_C - d_C$ additional events in the control arm.

Because there is a one-to-one mapping between $\deffh$ and $w$ under these assumptions, we can attempt to solve (\ref{enee_general}) numerically by finding the case weight $w$ that leads to a precision equal to that obtained with the hybrid model. In particular (see Appendix \ref{app:gen}), we can fit a model to the observed trial data (excluding external data) and find the common case weight $w$ that when applied to all control patients results in a precision equal to $\phhyb$ (giving a weight of 1 to all patients in the experimental arm).

As shown in our simulations, $\deffh$ estimated according to (\ref{enee_general}) can have large variability and bias. However, problematic estimates can be diagnosed by small derivatives of $\phrefk[\deff]$ prior to interpreting results. Simulations can help identify a threshold for the derivative prior to analysis, below which the solution would be ignored.

\section{Simulations \label{sim_section}}

\subsection{Exponential data}

We simulated interim analysis data for 1:1 randomized trials under four HRs between the experimental and control group. For each HR we simulated 1,000 datasets for a total of 4,000 simulated datasets. Each simulated trial contained between 60 and 100 total patients enrolled linearly at 2 patients per month. The interim was triggered after 33\% of the patients had an event, after which time all events were censored. For each simulated trial, we also simulated an external comparator cohort coming from the same distribution as the trial control and restricted follow-up time to the maximum follow-up in the trial to ensure comparability of endpoints. Please see Appendix \ref{app:sim} for details.

For each generated dataset, we fit an exponential model to the trial data excluding external data (reference model). We also fit an exponential model to the pooled external and trial data (representing the hybrid model). We used (\ref{ENEE_exact}) to estimate the number of events in the external comparator. To facilitate comparison, we used the EHSS approximation in terms of number of events instead of number of patients, given by
\begin{equation}
  \text{EHSS}_d \approx (d_C + d_E) \left( \frac{\phhyb}{\phref} - 1 \right).
  \label{ehssd}
\end{equation}
Because the external data come from the same generating distribution as the trial controls, the number of events in the external comparator represents the truth. All models were fit using the \verb+survival+ package \citep{survival2015} for R \citep{R2019}.

\begin{figure}[H]
  \centering
  \includegraphics[scale = 0.125]{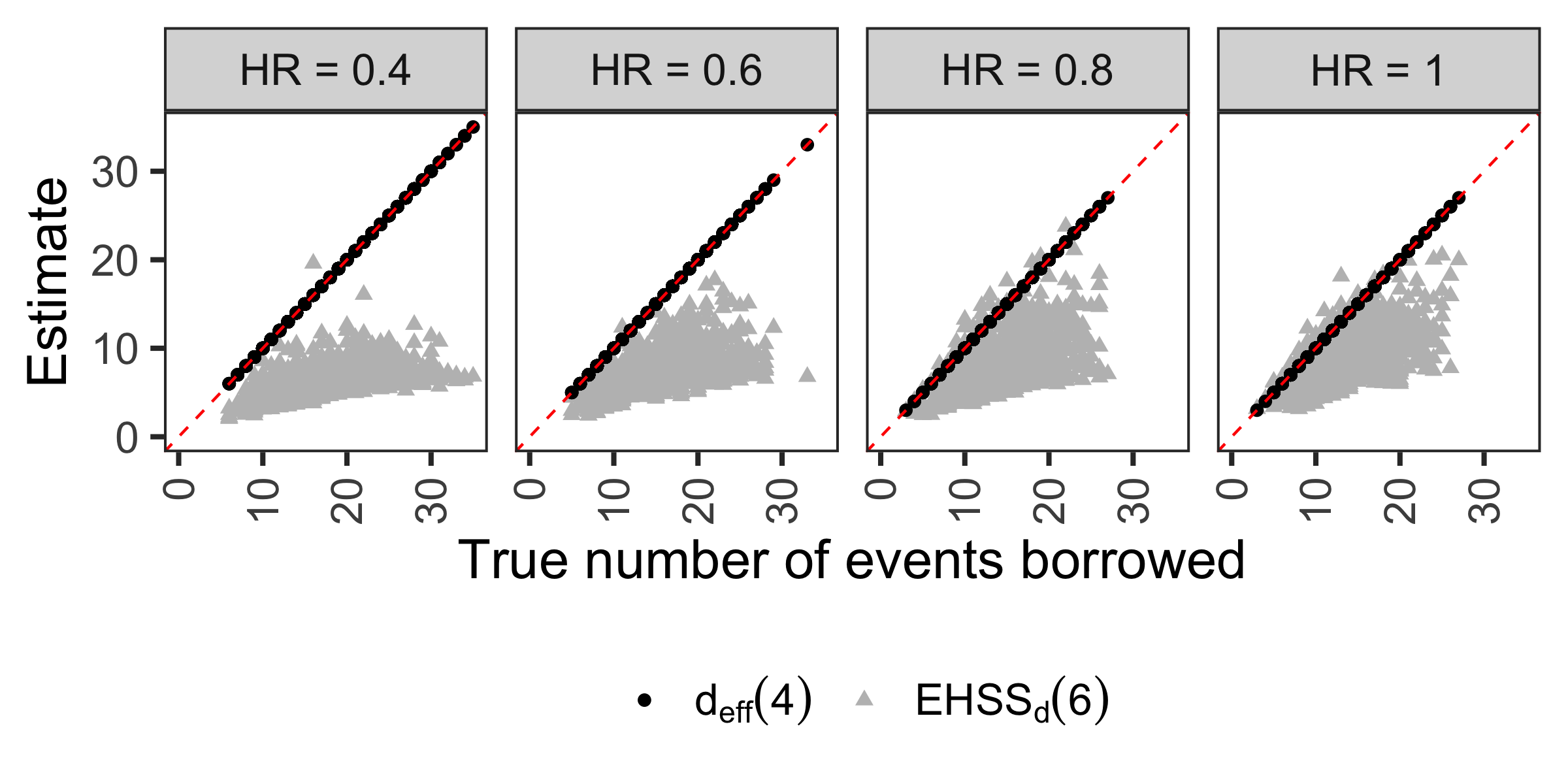}
  \caption{Simulation results with exponential data and models. The panels are split by the hazard ratio (HR) between experimental and control groups, and the diagonal dashed red lines have an intercept of 0 and a slope of 1 and indicate agreement. 1,000 datasets were generated for each HR.}
  \label{EHSS_ENEE_exponential_exp}
\end{figure}

Figure \ref{EHSS_ENEE_exponential_exp} shows simulation results with exponential data and models. As expected, the results for $\deffh$ estimated by (\ref{ENEE_exact}) are exact, whereas $\text{EHSS}_d$ (\ref{ehssd}) is a linear approximation that tends to underestimate the number of events borrowed in these simulations.

\subsection{Weibull data \label{sec:weib_sim}}

We also fit Cox models to simulated Weibull data, again using the \verb+survival+ package \citep{survival2015} for R \citep{R2019} to fit all models. The Weibull simulations were the same as the exponential simulations with the addition of a shape parameter that was set to 1.15 for an increasing hazard over time (see appendices for details). Of the 4,000 simulations, we deemed 1,239 estimates of $\deffh$ (obtained via (\ref{enee_general})) to be unstable as diagnosed by $ \partial \phrefk[\deff] / \partial \deff < \exp(- 3)$, where the derivative at $\deff = \deffh$ was approximated with a central finite difference. The root finding algorithm failed for an additional 67 simulations, which we also considered unstable, for a total of 1,306 unstable estimates.

The cutoff of $ \partial \phrefk[\deff] / \partial \deff < \exp(- 3)$ was chosen to exclude highly implausible values in this particular simulation. If this estimation procedure for $\deffh$ were used in practice, we would recommend that a cutoff value be chosen prior to the interim analysis based on simulations specific to the trial setting. If the derivative falls below the chosen \textit{a priori} cutoff at interim, then we would recommend using other methods to estimate $\deffh$.

\begin{figure}[H]
  \centering
  \includegraphics[scale = 0.125]{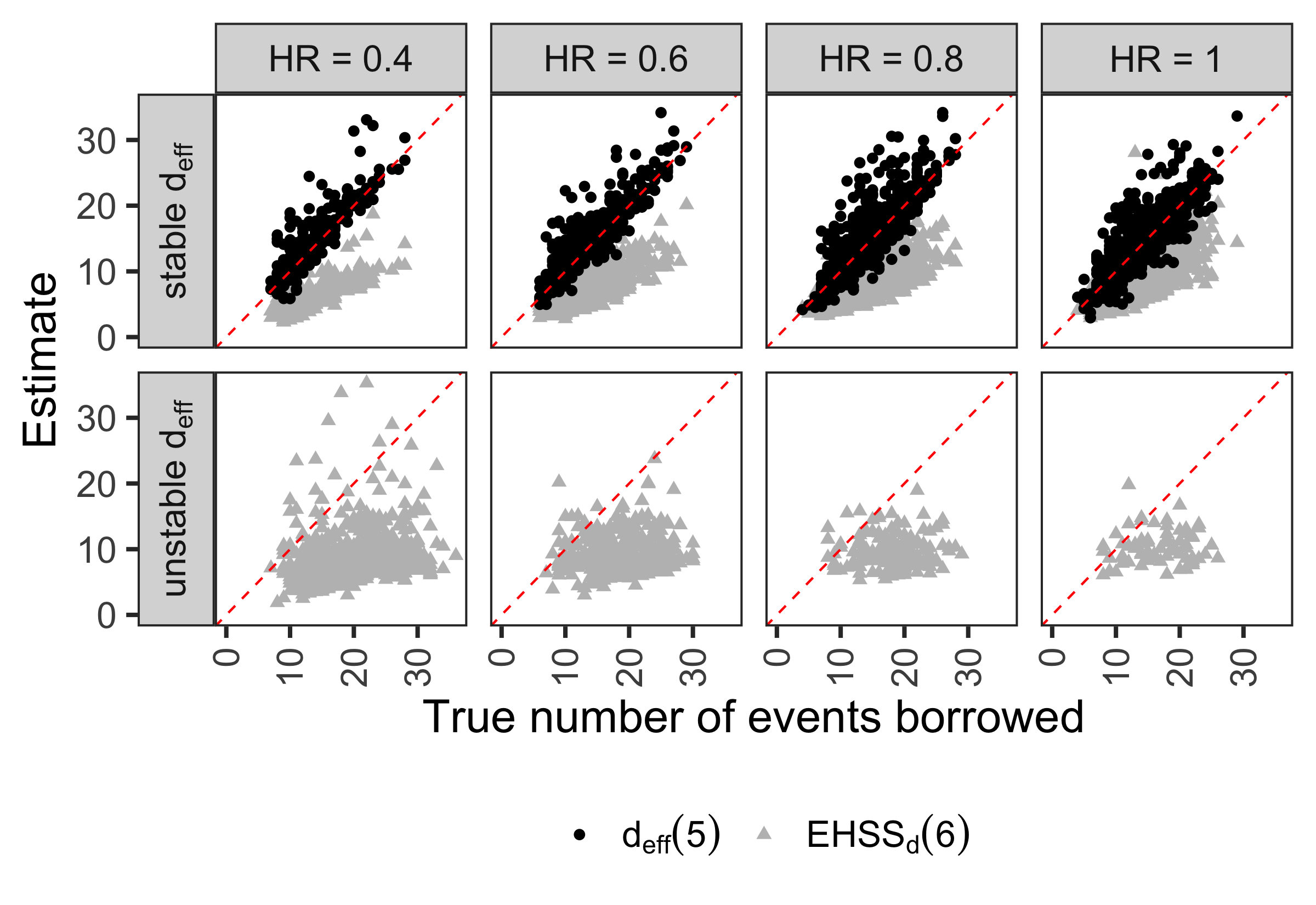}
  \caption{Simulation results with Weibull data and Cox model. The panels are split by the hazard ratio (HR) between experimental and control groups and whether $\deffh$ was deemed stable. The diagonal dashed red lines have an intercept of 0 and a slope of 1 and indicate agreement. $\deffh$ results that were deemed unstable have been removed (1,306 removed), but the corresponding $\text{EHSS}_d$ estimates are shown in lower panels. 1,000 datasets were generated for each HR.}
  \label{EHSS_ENEE_weibull_cox}
\end{figure}

Figure \ref{EHSS_ENEE_weibull_cox} shows the results with unstable $\deffh$ removed but with the corresponding $\text{EHSS}_d$ estimates shown in the lower panels. In practice, we can use the derivative diagnostic to assess the stability of $\deffh$ prior to interpreting results, but the same is not true of $\text{EHSS}_d$, which is why we retain all $\text{EHSS}_d$ estimates.

Table \ref{weibull_results_table} shows the empirical bias and standard deviation. The generalized estimate of $\deffh$ has lower bias and standard deviation than the linear approximation when restricting to estimates for which the root finding algorithm succeeded and $ \partial \phrefk[\deff] / \partial \deff \ge \exp(- 3)$, but as noted above, this excludes a large proportion of the simulation results (32.7\%). The full simulation results are shown in Appendix \ref{app:full_sim}, which highlight the importance of using diagnostics prior to interpreting results.

We also note that the linear approximation has higher standard deviation when the corresponding $\deffh$ is unstable than when it is stable. This suggests that the diagnostic for $\deffh$ may be useful even if another approach is ultimately used to obtain an estimate of the number of events effectively borrowed.

\begin{table}[H]
  \centering
  \caption{Empirical bias and standard deviation from Weibull simulations. Unstable $\deffh$ were removed but results for the corresponding $\text{EHSS}_d$ are shown in the rows labeled as ``Unstable.''}
  \begin{tabular}{cccccc}
    \toprule
    $\deffh$ Stability & HR & Method & \# Sims & Bias & St. Dev \\
    \midrule
    Stable   & 0.4 & $\deffh$ (\ref{enee_general})  & 301 &  0.876& 2.28 \\
    Stable   & 0.4 & $\text{EHSS}_d$ (\ref{ehssd}) & 301 & -7.27 & 2.82 \\
    Stable   & 0.6 & $\deffh$ (\ref{enee_general})  & 631 &  0.687& 2.01 \\
    Stable   & 0.6 & $\text{EHSS}_d$ (\ref{ehssd}) & 631 & -6.71 & 3.00 \\
    Stable   & 0.8 & $\deffh$ (\ref{enee_general})  & 836 &  0.571& 2.50 \\
    Stable   & 0.8 & $\text{EHSS}_d$ (\ref{ehssd}) & 836 & -5.82 & 3.39 \\
    Stable   & 1   & $\deffh$ (\ref{enee_general})  & 926 &  0.171& 2.39 \\
    Stable   & 1   & $\text{EHSS}_d$ (\ref{ehssd}) & 926 & -4.64 & 3.27 \\
    \midrule
    Unstable & 0.4 & $\text{EHSS}_d$ (\ref{ehssd}) & 699 &-12.0  & 6.09 \\
    Unstable & 0.6 & $\text{EHSS}_d$ (\ref{ehssd}) & 369 & -9.78 & 5.56 \\
    Unstable & 0.8 & $\text{EHSS}_d$ (\ref{ehssd}) & 164 & -8.49 & 5.00 \\
    Unstable & 1   & $\text{EHSS}_d$ (\ref{ehssd}) &  74 & -6.45 & 5.11 \\
    \bottomrule
  \end{tabular}
  \label{weibull_results_table}
\end{table}

\section{Discussion}

Adaptive trial designs using external data have the promise to improve the efficiency and feasibility of randomized controlled trials in settings where enrollment is challenging. The amount of data effectively borrowed from external data is a key piece of input to calculate changes to the randomization ratio, but may be challenging to quantify accurately outside of unadjusted exponential models.

We hypothesize that this difficulty is because in general, the estimated precision is not solely a function of the number of events. For example, in general the precision of a Cox regression is a function of both the number and relative timing of events across groups, as seen by inspecting the form of the information matrix. Consequently, up-weighting the trial controls to effectively create more events can only represent a narrow range of potentially observable outcomes had additional controls been enrolled in the trial, particularly with small sample sizes. As seen in Figure \ref{EHSS_ENEE_weibull_cox} and Table \ref{weibull_results_table}, diagnostics for our generalized solution (\ref{enee_general}) can also identify situations in which the standard linear approximation has higher bias and variance, suggesting that some scenarios may be inherently challenging.

\citet{morita2008} minimized the distance between two information matrices to estimate the effective sample size of a Bayesian prior distribution. As we demonstrate, minimizing the distance between two variances (or equivalently, two precisions) can also be used to determine effective sample size outside of a Bayesian context, and can be applied to frequentist and semi-parametric models such as Cox regression. This may help to enable the use of a wider variety of methods in adaptive hybrid controlled trials.

\section{Conclusion}

For unadjusted exponential models, we recommend using the exact solution given by (\ref{ENEE_exact}) to estimate the number of external events effectively borrowed. For all other situations, we recommend conducting trial specific simulations to assess the behavior of different methods, such as the standard linear approximation and proposed generalized solution. The simulations can help guide the choice of method and diagnostic, highlight limitations, and improve the interpretation of the estimated amount of borrowed information.

\section*{Supplementary material}
\text{R} code for the simulations is available at \\
\url{https://github.com/flatironhealth/code-for-enee-paper}.

\section*{Acknowledgments}
We would like to thank Brian Hobbs, Daniel Backenroth, Meghna Samant, and Somnath Sarkar for helpful discussions.

\bibliographystyle{apalike}
\bibliography{enee_refs.bib}

\appendix
\appendixpage

\section{Generalized solution \label{app:gen}}

As noted in the Section \ref{sec:gen}, for a given case weight $w$, the effective number of additional events in the control arm of the trial is $\deff = w d_C - d_C = d_C(w - 1)$. We also have the constraint that $w = \deff / d_C + 1 > 0$, which implies $\deff > -d_C$.

For a given $\deff$, we set $w = \deff / d_C + 1$ and fit the reference model to the trial data where patient $i$ gets weight 
\begin{equation*}
  w_i = 
  \begin{cases}
    w \text{ if patient $i$ is in control arm}\\
    1 \text{ if patient $i$ is in experimental arm.}
  \end{cases}
\end{equation*}

Let $\psi(\deff) = \phrefk[\deff] - \phhyb$. We numerically solve for $\deffh$ such that $\psi(\deffh) = 0$. In the simulations, we searched for a solution over the interval $(-d_C + 0.001, 1000)$. We then approximated $\partial \phrefk[\deff] / \partial \deff$ at $\deff = \deffh$ by
\begin{equation*}
  \left. \frac{\partial \phrefk[\deff]}{\partial \deff} \right|_{\deff = \deffh} \approx \frac{\phrefk[\deffh + \epsilon] - \phrefk[\deffh - \epsilon]}{2 \epsilon}.
\end{equation*}
In the simulations, we set $\epsilon = 0.0001$.

\section{Details of simulations \label{app:sim}}

\textbf{Exponential data:} For each of four hazard ratios between experimental and control groups (0.4, 0.6, 0.8, and 1), we conducted 1,000 simulations for a total of 4,000 simulated datasets. For each simulation, we randomly selected the size of the trial uniformly between 60 and 100, and a censoring probability uniformly between 0.05 and 0.1. We triggered the interim after 33\% of patients had an observed event (censoring all events occurring after this time). We also required that at least 10 event had occurred in each arm before triggering the interim to prevent pathological simulations.

We generated trial data assuming a linear enrollment of 2 patients per month with a 1:1 randomization ratio and exponential outcomes with a hazard rate of $1/12$ in the control arm (mean of 12 months-to-event). We then randomly selected the number of external patients uniformly between 10 and the number of control patients in the trial who had enrolled by the time the interim was conducted, and simulated exponential outcomes with the same hazard and censoring rate as the trial controls. External patients were censored at the max follow-up time in the trial to ensure comparability of endpoints.\\

\textbf{Weibull data:} For the Weibull simulations, we followed the same approach as above and set the shape parameter to 1.15 for an increasing hazard over time. For the Weibull simulations, the expected censoring rate was not exactly as specified.

\section{Full results of Weibull simulation \label{app:full_sim}}

\begin{figure}[H]
  \centering
  \includegraphics[scale = 0.125]{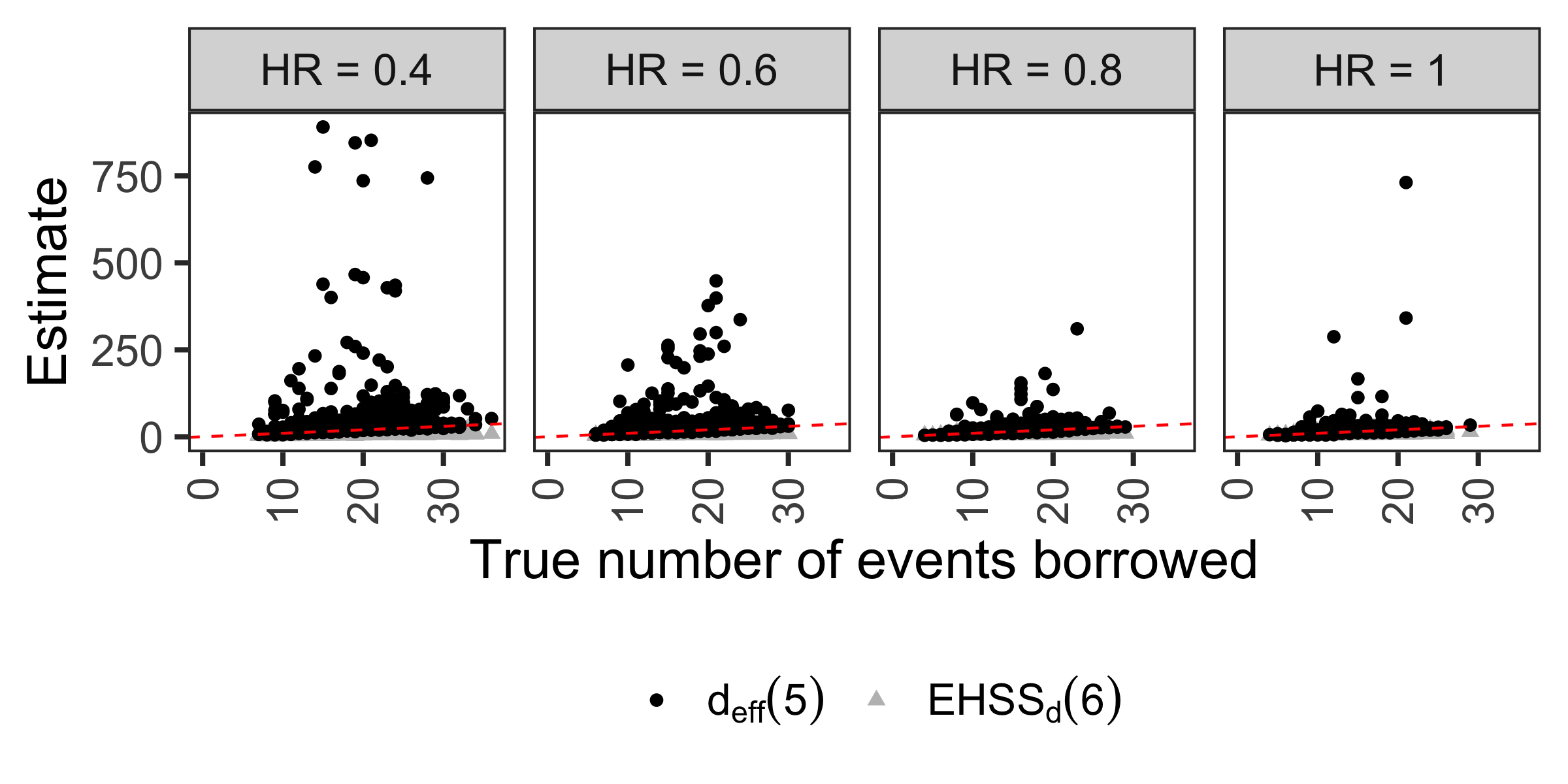}
  \caption{Simulation results with Weibull data and Cox models without removing unstable results. The panels are split by the hazard ratio (HR) between experimental and control groups and the diagonal dashed red lines have an intercept of 0 and a slope of 1 and indicate agreement. 1,000 datasets were generated for each HR.}
  \label{EHSS_ENEE_weibull_cox_all}
\end{figure}

As noted in Section \ref{sec:weib_sim}, 1,306 simulations had unstable $\deffh$ estimates (1,239 had small $\partial \phrefk[\deff] / \partial \deff$ and 67 had solutions outside of $(-d_C + 0.001, 1000)$). In practice, we would recommend defining a cutoff for $\partial \phrefk[\deff] / \partial \deff$ \textit{a priori} based on simulations tailored to the trial at hand, and only using the results of the generalized solution if $\partial \phrefk[\deff] / \partial \deff$ is larger than the predefined threshold. It could also be helpful to set a reasonable upper bound over which to search with the root-finding algorithm. To provide insight into the problems that could arise if diagnostics and reasonable upper limits are not taken into account, Figure \ref{EHSS_ENEE_weibull_cox_all} shows all simulation results. As can be seen in Figure \ref{EHSS_ENEE_weibull_cox_all}, in these simulation scenarios the instability can lead to enormous errors, highlighting the importance of using diagnostics prior to interpreting results.

\end{document}